\documentclass[twocolumn]{aastex63}

\newcommand{\kms}{${\rm km\,s^{-1}}$}

\newcommand{\cc}{$^{13}$CO}
\newcommand{\ccc}{$^{13}$C$^{18}$O}
\newcommand{\ce}{C$^{18}$O}
\newcommand{\cseven}{C$^{17}$O}

\revised{}
\shorttitle{Volatile carbon evolution}
\shortauthors{Zhang et al.}
\graphicspath{{./}{figures/}}

\begin{document}

\title{Rapid Evolution of Volatile CO from the Protostellar Disk Stage to the Protoplanetary Disk Stage}

\correspondingauthor{Ke Zhang}
\email{kezhang@umich.edu}

\author[0000-0002-0661-7517]{Ke Zhang}
\affiliation{Department of Astronomy, University of Michigan, 
323 West Hall, 1085 S. University Avenue, 
Ann Arbor, MI 48109, USA}
\affiliation{Hubble Fellow}

\author[0000-0002-6429-9457]{Kamber Schwarz}
\affiliation{Lunar and Planetary Laboratory, University of Arizona,
1629 E. University Boulevard,
Tucson, AZ 85721, USA}
\affiliation{Sagan Fellow}

\author[0000-0003-4179-6394]{Edwin A. Bergin}
\affiliation{Department of Astronomy, University of Michigan,
323 West Hall, 1085 S. University Avenue,
Ann Arbor, MI 48109, USA}



\begin{abstract}

Recent observations show that the CO gas abundance, relative to H$_2$, in many 1-10\,Myr old protoplanetary disks may be heavily depleted, by a factor of 10-100 compared to the canonical interstellar medium value of 10$^{-4}$. When and how this depletion happens can significantly affect compositions of planetesimals and atmospheres of giant planets. It is therefore important to constrain if the depletion occurs already at the earliest protostellar disk stage.
Here we present spatially resolved observations of \ce, \cseven, and \ccc~$J$=2-1 lines in three protostellar disks. We show that the \ce~line emits from both the disk and the inner envelope, while \cseven~and \ccc~lines are consistent with a disk origin. The line ratios indicate that both \ce~and \cseven~lines are optically thick in the disk region, and only \ccc~line is optically thin.  The line profiles of the \ccc~emissions are best reproduced by Keplerian gaseous disks at similar sizes as their mm-continuum emissions, suggesting small radial separations between the gas and mm-sized grains in these disks, in contrast to the large separation commonly seen in protoplanetary disks. Assuming a gas-to-dust ratio of 100, we find that the CO gas abundance in these protostellar disks are consistent with the ISM abundance within a factor of 2, nearly one order of magnitude higher than the average value of 1-10\,Myr old disks. These results suggest that there is a fast, $\sim$1~Myr, evolution of the abundance of CO gas from the protostellar disk stage to the protoplanetary disk stage.

\end{abstract}

\keywords{astrochemistry --- circumstellar matter --- molecular data ---protoplanetary disks --- protostars }


\section{Introduction} \label{sec:intro}
Circumstellar disks around young stars set the stage for planet formation \citep{williams11}.  The chemical composition of natal disks may change dramatically from the formation of disks in embedded protostars to the time of disk gas dispersal several Myr later. These changes will profoundly affect the compositions of planets formed at different disk locations and timescales.

One particularly surprising discovery regarding the composition of disk material is that the abundance of volatile gas-phase CO in many protoplanetary disks (1-10\,Myrs old, Class II) may be one or two orders of magnitude lower than the canonical value. First, significant CO depletion (by a factor of 5-100) was found in all three protoplanetary disks that have independent gas mass measurements from HD (1-0)\,line fluxes \citep{Favre13, McClure16, Schwarz16, Kama16, Trapman17, Zhang17}. Moreover, surprisingly weak CO emissions are seen in the majority of Class II disks in nearby star formation regions, even after correcting the effects of CO freeze-out and isotope-selective photodissociation \citep{Ansdell16, Miotello16, Long17, Zhang19}. This weak CO problem can be explained either by more than one order of magnitude of CO depletion, or by a rapid gas dissipation. But a gas-rich condition (gas-to-dust ratio $\sim$100) is required to explain mass accretion rates of these disks \citep{Manara16} and to match the observed outer edges of mm-dust emission with that predicted by dust drifting models \citep{Powell19}. 
These results all point to significant CO depletion in many Class II disks.

It is still unclear when and how the CO gas depletion happens. Theoretical works suggest that the abundance of CO gas in the disk atmosphere can be reduced either by chemical processes that turn CO into other less volatile carbon-species \citep{bergin14,Eistrup16,schwarz18,Bosman18,Schwarz19a}, or by dust growth  processes  that  sequester  CO  ice  into  the  disk midplane \citep{xu17, Krijt18}. Current models require 1\,Myr or longer time to reduce the CO gas abundance by a factor of ten under typical conditions in Class II disks.

It is of particular interest to determine whether CO gas is already depleted at the youngest disk stage (Class 0/I), $<$1\,Myr, as growing number of observations suggest the first steps of planet formation may already start at the protostellar disk stage \citep{alma15, zhang15b, Harsono18}. Rapid evolution of CO gas abundance will lead to significant variations in carbon inventory of planetesimals and atmospheres of giant planets formed at different timescales. 

Measuring the abundance of CO gas in protostellar disks is challenging due to line confusion from the surrounding envelopes and outflows. There have been few observational constraints of CO abundances attributed to the disk alone, and existing ones gave conflicting results. On the inner envelope scale ($\sim$1000\,AU), \ce~(2-1) images of four Class 0 sources suggest significant CO depletion \citep{Anderl16}, but Herschel-HIFI observation of high $J$ (up to $J$=10) CO transitions of deeply embedded low-mass Class 0/I systems do not show depletion \citep{Yildiz13}. 
Scaling down to the disk region, fluxes of \ce\,(2-1) line from the HL Tau and V2775 Ori disks suggest gas-to-dust mass ratios of $<$10, which potentially implies a factor of 10 depletion in CO abundance \citep{Yen17, Zurlo17}.  However, even \ce~lines can be optically thick in protostellar disks \citep{vantHoff18}. A more robust way is to employ rarer CO isotopologue lines (e.g., \cseven, \ccc,$^{13}$C$^{17}$O ) that have smaller optical depths \citep{Zhang17, ArturdelaVillarmois18, Booth19,Booth20}.

Here we report spatially resolved observations of \ce/\cseven/\ccc~(2-1) lines towards three young protostellar disks ($<$1\,Myr). Our goal is to measure the CO gas abundance in the disk area and to test whether it evolves significantly from the protostellar disk stage to the protoplanetary disk stage.

\section{Observations} \label{sec:obs}

\begin{deluxetable*}{cccccccccl}[!t]
\tablecaption{Source information \label{tab:source}}
\tablehead{
 \colhead{Source} &  \colhead{M$_\star$} &  \colhead{T$_{\rm eff}$} &  \colhead{L$_{\rm bol}$} &  \colhead{M$_{\rm disk}$} &  \colhead{Distance} &  \colhead{incl} &  \colhead{disk PA} &  \colhead{outflow PA} &  \colhead{Type
} \\
 \colhead{} &  \colhead{(M$_{\odot})$} &  \colhead{(K)} &  \colhead{(L$_{\odot}$)} &  \colhead{(M$_{\odot})$} &  \colhead{(pc)} &  \colhead{(deg)} &  \colhead{(deg)} &  \colhead{(deg)} &  \colhead{
} 
}
\startdata
TMC1A & 0.53 & 4000 & 2.5 & 0.05 & 140 & 55 & 67 & -17 & Class I\\
HL Tau & 1.7 & 4000 & 11 & 0.13 & 140 & 46.2 & 138.2 & 50 & Class I/II\\
DG Tau & 0.7 & 4775 & 6.4 & 0.02 & 121 & 41 & 120 & 226 & Class I/II\\
\enddata
\end{deluxetable*}

Our protostellar sample consist of three sources: the TMC1A, HL Tau, and DG Tau systems, all with known large Keplerian rotating disks ($>$100\,au, \citealt{Isella10,alma15, Aso15, Harsono18}). We chose these large protostellar disks so that these disks and their inner envelopes can be spatially resolved with NOEMA C or D configurations. 

 The observations were carried out with the NOEMA interferometer between November 2017 and April 2019. The total on-source integration time was 7.3, 5.4, and 5.3 hours for HL Tau, TMC1A, and DG Tau, respectively. All observations used the new wide-band correlator PolyFix with an instantaneous dual-polarization coverage of 15.5\,GHz bandwidth at a fixed resolution of 2000\,kHz. In addition, higher spectral resolution chunks were set at the line centers of \cc, \cseven, and \ccc~(2-1) with a resolution of 65\,kHz. The baseline lengths were between 24 to 704\,m. Nearby quasars were observed between science targets to calibrate the complex antenna gains. The absolute flux calibrations were obtained by observing MWC 349 and/or LKHA 101. Data calibration and imaging were done using the GILDAS software. We re-binned data to a channel width of 0.5\,\kms~to enhance signal-to-noise ratios. After uniform weighting, the synthesized beams are 1\farcs4$\times$1\farcs2~for TMC1A and DG Tau, and 2\farcs3$\times$1\farcs7~for HL Tau. The noise levels are 4.6\,mJy beam$^{-1}$ in individual channels. The absolute flux uncertainty is expected to be 15-20\% and all 1.3\,mm continuum fluxes were consistent with literature values within 15\%.

\begin{figure*}[ht]
\epsscale{1.2}
\vspace{-1.4cm}
\plotone{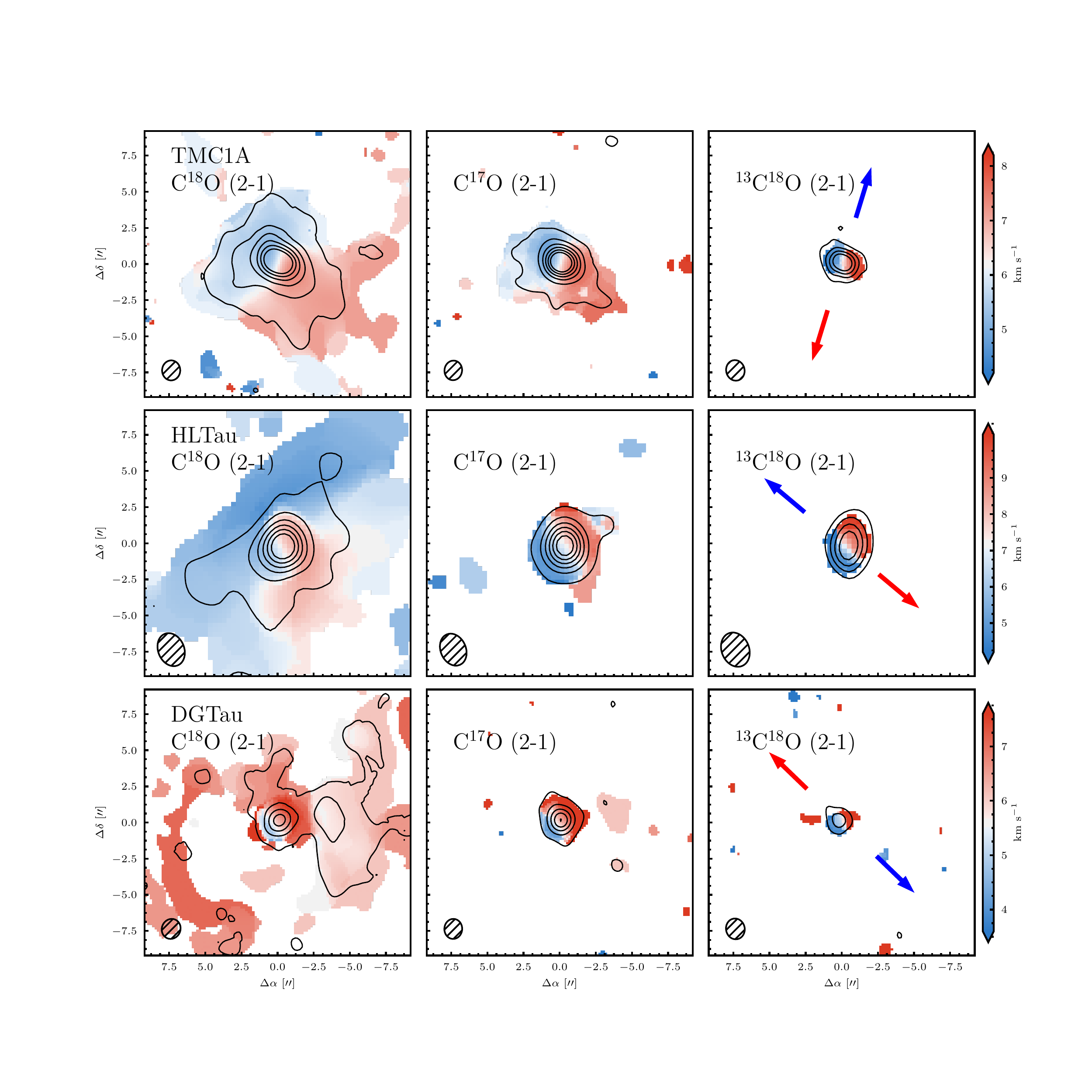}
\vspace{-1.4cm}
\caption{Observations of CO isotopologue (2-1) lines towards TMC1A, HL Tau, and DG Tau systems. The integrated line intensity maps are plotted in contours and the velocity maps are in color. For all lines, contours start at the 3$\sigma$ level on the integrated images, and increase with 3$\sigma$ interval for \ccc~, and with 10$\sigma$ interval for the \ce~and \cseven~lines. The position angles of known outflows/jets of the three systems are indicated in \ccc~panels. \label{fig:co_images}}
\vspace{0.6cm}
\end{figure*}

 The three isotopologue CO lines were detected in all targets. Figure~\ref{fig:co_images} shows the integrated line emission and velocity maps of the CO lines. The \ce~line emissions are generally from 10-15\arcsec~wide regions and show complicated velocity structures from both disks and surrounding inner envelopes. The \cseven~and \ccc~line emissions are much more compact and centered on the area traced by the dust thermal continuum. Their velocity maps are consistent with position angles of the continuum emission and perpendicular to known outflow/jet in each system. The compact emitting regions and velocity maps suggest that \cseven~and \ccc~line emissions are primarily arising from within the protostellar disk. In Table~\ref{tab:line_flux}, we list line fluxes integrated from a 5\arcsec$\times$5$\arcsec$ box region. The integrated fluxes of \ce~and \cseven~lines increase with box sizes, but the fluxes of \ccc~line do not change as the emissions are compact. 
 
 To estimate the optical depth of different lines, we measured ratios of surface brightness temperature (T$_b$) among three lines at regions where \ccc~emission is greater than 3$\sigma$. We found that the T$_b$(\ce)/T$_b$(\cseven)$\sim$1 and T$_b$(\cseven)/T$_b$(\ccc)~between 5-12. In all cases, line ratios are less than the local ISM abundance ratios of $^{18}$O/$^{17}$O=3.6 and $^{12}$C/$^{13}$C=69 \citep{wilson99}. These lower ratios suggest that \ce~is highly optically thick and \cseven~is marginally optically thick ($\tau\sim$2-3) within these disks. Therefore, we use optically thin \ccc~line spectra to measure the CO gas masses.

\begin{deluxetable*}{cccccc}
\tablecaption{Line fluxes and best-fit CO abundance\label{tab:line_flux}}
\tablehead{\colhead{Source} & 
 \colhead{\ce\,(2-1)} &  \colhead{\cseven\,(2-1)} &  \colhead{\ccc\,(2-1)} &  \colhead{F$_{\rm 1.3\,mm}$} & \colhead{n$_{\rm CO}$/n$_{\rm H_2}$} \\
 \colhead{}&\colhead{(mJy.\kms)} &  \colhead{(mJy.\kms)} &  \colhead{(mJy.\kms)} & \colhead{(mJy)}  &
}
\startdata
TMC1A & 4954$\pm$50 & 3307$\pm$48 & 306$\pm$30 & 183$\pm$8 & 2.1$^{+0.1}_{-0.2}\times10^{-4}$\\
HL Tau & 3195$\pm$69 & 3357$\pm$75 & 329$\pm$37 & 752$\pm$10  & 6.7$^{+0.6}_{-0.5}\times10^{-5}$ \\ 
DG Tau & 872$\pm$33 & 781$\pm$36 & 107$\pm$27 & 351$\pm$11 & 4.6$^{+0.4}_{-0.4}\times10^{-5}$\\
\enddata
\end{deluxetable*}

\section{Methods} \label{sec:methods}

\subsection{Model setup}
\begin{figure*}[!htbp]
\plotone{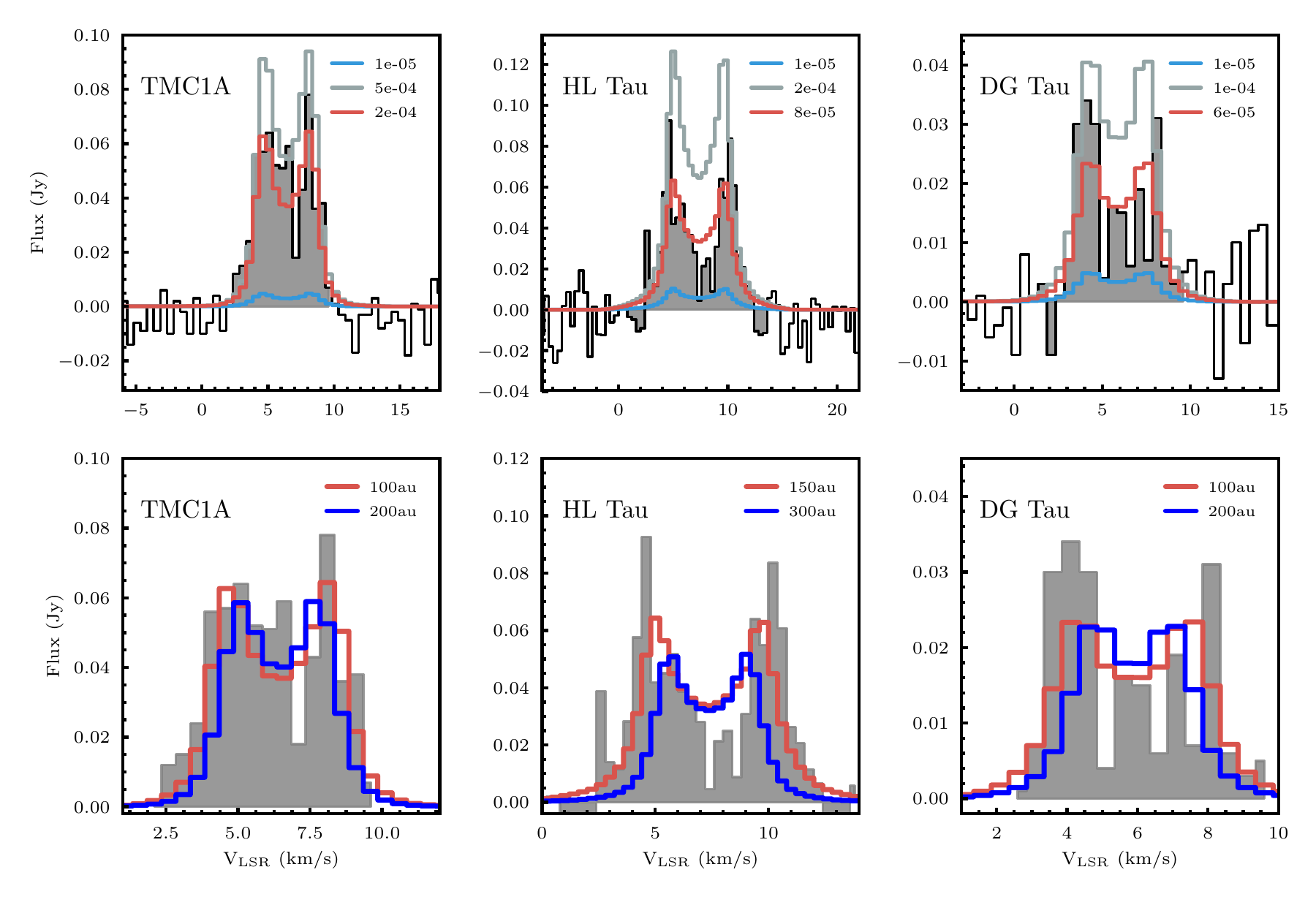}
\caption{The integrated line profiles of observed \ccc~(2-1) line (grey) compared with model spectra. Upper row: models of different CO abundances. Models of best-fit abundances, twice of the best-fit abundances, and a CO abundance of 10$^{-5}$ are plotted in red, light grey, and blue, respectively. These results show that the CO gas abundances in the three disks should be 5-20 times higher than the 10$^{-5}$ value seen in many Class II disks. Bottom row: Models of the same disk sizes of gas and mm-dust are in red, and models of gas disks twice the size of dust disks are in blue. \label{fig:models}}
\vspace{0.6cm}
\end{figure*}

Assuming the \ccc~line emission is dominantly from the protostellar disk, we adopt a parameterized disk model to derive the CO gas abundances in these disks. The global surface density is from the self-similar solution of a viscously evolving disk \citep{Lynden-Bell74}:
\begin{equation}
\Sigma (R) = \Sigma_c \Big(\frac{R}{R_c}\Big)^{-\gamma} {\rm exp} \Big[ -\Big(\frac{R}{R_c}\Big)^{2-\gamma}\Big], 
\end{equation}
 where $\Sigma_c$ is the surface density at the characteristic radius R$_c$, and $\gamma$ is the gas surface density exponent. 

The disk model includes three mass components -- gas, a population of small dust, and a population of large dust. The small and large dust populations are assumed to be 5\% and 95\% of the total solid mass, respectively. 
Previous observations indicated that significant dust growth and settling may already occur in Class 0/I disks \citep{Pinte16,Harsono18}. The two dust populations are employed to mimic these effects. 
The gas and the small grain population are assumed to be spatially coupled and characterized by a scale height distribution of H$_g$(r)=H$_0(r/r_0)^{1.2}$. The large grain population is more settled and has a scale height reduced by a factor of $\chi$, i.e., H = $\chi$H$_g$ with $\chi<$1. We use $\chi$=0.1 for the HL Tau disk, because observations showed that mm-sized dust in this disk is extremely settled \citep{Pinte16}. For other two disks, we adopt $\chi$=0.2, a value widely used in literature. Both dust populations are assumed to follow the power-law size distribution of $n(a)\propto a^{\rm -3.5}$ with a minimum grain size of a$_{\rm min}$=0.005\,$\mu$m and a$_{\rm max}$=1\,$\mu$m for the small grain population and a$_{\rm max}$=1\,mm for the large grain population.  

 The disk size of the large ($\sim$mm-sized) dust population can be constrained by spatially resolved (sub)mm continuum observations, but disk sizes of gas and small grains are more uncertain. In this regard, many Class II disks show that their CO gas disk is much larger than the continuum emission \citep{ansdell18}, which is usually attributed to inward radial drift of mm-sized grains \citep{Birnstiel14, Trapman19}. Here we start with models that all three components have the same radial surface density profile. We will show later that larger gas disks become inconsistent with our observations of CO line profiles.

For the CO abundance structure, we use a setup similar to \citet{williams14}. The CO gas abundance is assumed to be a constant across the whole disk region except for the regions cooler than the CO freeze-out temperature (20\,K) and above the photo-dissociation layer. This layer is referred to as the ``warm molecular layer'' as described by \citet{Aikawa02}. We note that even within this layer the CO gas abundance may vary with location in the disk due to chemical and physical processes \citep{Zhang19}, but our observations do not have sufficient spatial resolution to constrain these variations. In our model photo-dissociation is assumed to happen above the N$_{\rm H_{2}}\sim10^{21}$\,cm$^{-2}$ layer. 

We use the radiative transfer code RADMC3D to calculate the thermal structure of the disk \citep{radmc3d}. The stellar mass and luminosity are adopted from Table~\ref{tab:source}. We do not consider viscus heating as it is expected to mostly affect the inner 10\,au \citep{Harsono15}. Gas and dust temperatures are assumed to be the same, as expected for the \ccc~line emitting region \citep{Zhang17}.  We first adopt dust masses from literature and then adjust the masses to match with 1.3\,mm continuum fluxes. All of the adjustments are within 20\% of the initial values. The gas-to-dust mass ratio is assumed to be 100, which is expected for protostellar disks and we will discuss the uncertainty of the total disk mass later. We then run a grid of models starting from a CO-to-H$_2$ abundance of 10$^{-6}$ to 10$^{-3}$, increasing by a factor of 1.1 at each step. We then compare the grid of models with the observed spectra of \ccc~line to find the best-fit CO abundance for each disk. These results are used to find an initial best-fit CO abundance, and then a finer grid with an increasing step of 1.02 is computed around the initial best-fit value to calculate uncertainties. 

We also test a second group of models by setting the gas disk to be a factor of two larger than the dust disk. The choice of a factor of two is based on the typical size ratio of the CO gas to (sub)mm continuum disks seen in Class II disks \citep{ansdell18}. The goal is to test if the \ccc~line emission is from a much larger radial region than the (sub)mm continuum disk, due to the radial drift of dust, like seen in many Class II disks.

\subsection{Model Results}
Our first model group has the same disk sizes for both the gas and dust. Its best-fit CO abundances are presented in Figure~\ref{fig:models} and Table~\ref{tab:line_flux}. We find that the CO gas abundances in the warm molecular layer of the three protostellar disks are between 5$\times$10$^{-5}$ and 2$\times$10$^{-4}$, which is consistent with the canonical ISM value within a factor of 2. These abundances are nearly an order of magnitude higher than that of the majority of Class II disks. 

For the second group of models (the gas disk is a factor of two larger than the mm-dust disk), we find that the velocity separations of two peaks of \ccc~line profiles become too small to be consistent with observations (see Figure~\ref{fig:models}). This suggests that the sizes of the CO gas disks are comparable to the continuum disks, which is different from the cases of many Class II disks.

\subsection{Uncertainty of CO abundances}
Table~\ref{tab:line_flux} shows that the CO abundances in our best-fit models have uncertainties on the level of 10\% for the given gas masses. However, the absolute values depend on gas disk masses, and therefore the abundance uncertainties are dominated by uncertainties in gas masses. In our models, we derive gas masses based on the mm-continuum fluxes and assume a gas-to-dust mass ratio of 100, as expected for the very young ($<$1~Myr old) protostellar disks. As our dust masses only include masses from grains up to mm-size, we might underestimate the total solid masses and subsequently the gas masses. However, the total disk masses in our models are already 3-10\% of the masses of the central protostars (see Table~\ref{tab:source}). If the disk masses were ten times higher, then these disks would be gravitationally unstable and show large-scale spiral structures in their (sub)mm continuum emissions \citep{Perez16}. However, none of these disks show significant asymmetric substructures in their high-resolution ALMA continuum observations. It is therefore unlikely that the actual gas masses are ten times higher than our models, and the ISM level CO abundances derived here cannot be attributed to the uncertainty of gas mass alone. 

Detailed dust evolution models of one of our sample disks, the HL Tau disk, requires a gas-to-dust ratio of 50 to match with continuum observations \citep{Tapia19}. Using the line flux of $^{13}$C$^{17}$O (3-2) and a disk-averaged temperature of 25\,K, \citet{Booth20} derived a total gas mass of 0.2\,M$_\odot$ for the HL Tau disk, which is 1.5 times higher than the gas mass we use. Based on these independent estimations, we estimate that the uncertainties of our CO abundances are within a factor of 2.   

\subsection{Sample selection bias}

Our study of CO abundance in protostellar disks is a small sample and biased to sources with known large Keplerian disks. So far there is no complete survey of the size and mass distribution of Keplerian rotating disks around Class 0/I sources. But our sample is most likely among the largest and most massive protostellar disks, as they are the most easily observable. 

One way to evaluate the statistical significance of our results is to compare our sample with the most massive Class II disks. \citet{Ansdell16} studied 89 Class II disks in the Lupus region and showed that nine out of the top ten most massive disks (based on dust mass) have CO abundance depletion by a factor of $\ge$5 compared to the canonical value. Similarly, \citet{Long17} studied 93 Class II disks in the Chamaeleon I region and found that eight out of the ten most massive disks have a factor of $\ge$5 depletion in their CO gas abundance. We calculated a Kolmogorov$-$Smirnov test for the CO abundances of our sample and these of the top 10\% most massive disks in the Lupus and Chamaeleon regions. The result shows that there is only a 1.7\% probability that these two samples are drawn from the same distribution. Therefore, within the largest and most massive disk population, our results suggest that the CO gas abundance in protostellar disks is significantly higher (by a factor of 5-10) than that in protoplanetary disks.  

\section{Discussions} \label{sec:discussion}
\subsection{Rapid evolution of the CO gas abundance}
\label{sec:co_evolution}
We show that three protostellar disks ($\le$1\,Myr) have CO gas abundances consistent with the ISM abundance within a factor of two. As shown in Figure~\ref{fig:co_evolution}, these CO abundances are nearly one order of magnitude higher than the average values of protoplanetary disks. If these CO abundances are representative of protostellar disks $<$1\,Myr, it has two important implications: first, the CO gas depletion seen in 1-10\,Myr disks are results of processes which occurred inside disks rather than the infalling envelopes; second, the process is extremely efficient -- it depletes volatile CO gas by a factor of ten within 1\,Myr. The timescale of the CO depletion process therefore is comparable with the general timescale of planetesimal/planet formation \citep{kruijer14}. Whatever mechanism(s) drive the CO depletion, it can significantly affect the final compositions of planets. 

Our measurement of CO gas abundances in protostellar disks are one order of magnitude higher than those measured in the warm inner envelops of four Class 0 sources \citep{Anderl16}. One way to reconcile these results is that the low CO abundances in the envelopes are due to CO freezing out onto water ice or CO$_2$ ices, instead of pure CO ice. The sublimation temperatures of these mixed CO ices are 10-15\,K higher than that of the pure CO ice \citep{Cleeves14}.

Our measurements put tight constraints on the timescale of the CO depletion. Because the stellar photosphere of DG Tau is more exposed than for the other two sources, DG Tau is the most evolved source ($\sim$1\,Myr) in our sample. However, the majority of disks in Lupus and Cha I (2-3\,Myr) already show CO depletion by a factor of 10-100. This rapid change suggests that once the CO depletion process begins, it is a really efficient process. 

\subsection{CO depletion mechanisms}

The CO depletion in Class II disks is generally attributed to two types of processes: chemical processes that turn CO into less volatile molecules, and/or dust growth processes that sequester CO ice into the disk midplanes. In the following, we discuss implications of our results for both types of processes. 

Current chemical models indicate that an ISM level cosmic-ray ionization rate (10$^{-17}$ s$^{-1}$) is essential for the chemical processing of CO \citep{Reboussin15,Yu16, Eistrup18,Bosman18,schwarz18,Schwarz19a}. But it is still unclear if the cosmic-ray ionization rate can reach that high level in 1-10\,Myr old protoplanetary disks \citep{Cleeves15}. On the other hand, the cosmic-ray rate might be orders of magnitude higher in protostellar disks due to accretion shocks \citep{Padovani16}. 

Temperature is another important parameter for the efficiency of the CO chemical processing. Given sufficient ionization, chemical models show that the processing occurs most efficiently at 15-30\,K \citep{schwarz18,Bosman18}. This temperature range works the best because the main pathways of CO processing are via CO reactions with OH or H on grain surfaces to form CO$_2$ or CH$_3$OH generally.   The efficiency of these processes depends on both the amount of CO freeze-out and the speed of reactions on grain surface. Higher temperatures reduce the freeze-out of CO onto grain surfaces, while lower temperatures reduce the efficiency of some grain surface reactions. Depending on the disk structure, chemical processing can work in some regions of protostellar disks. 

\begin{figure*}[!htp]
\plotone{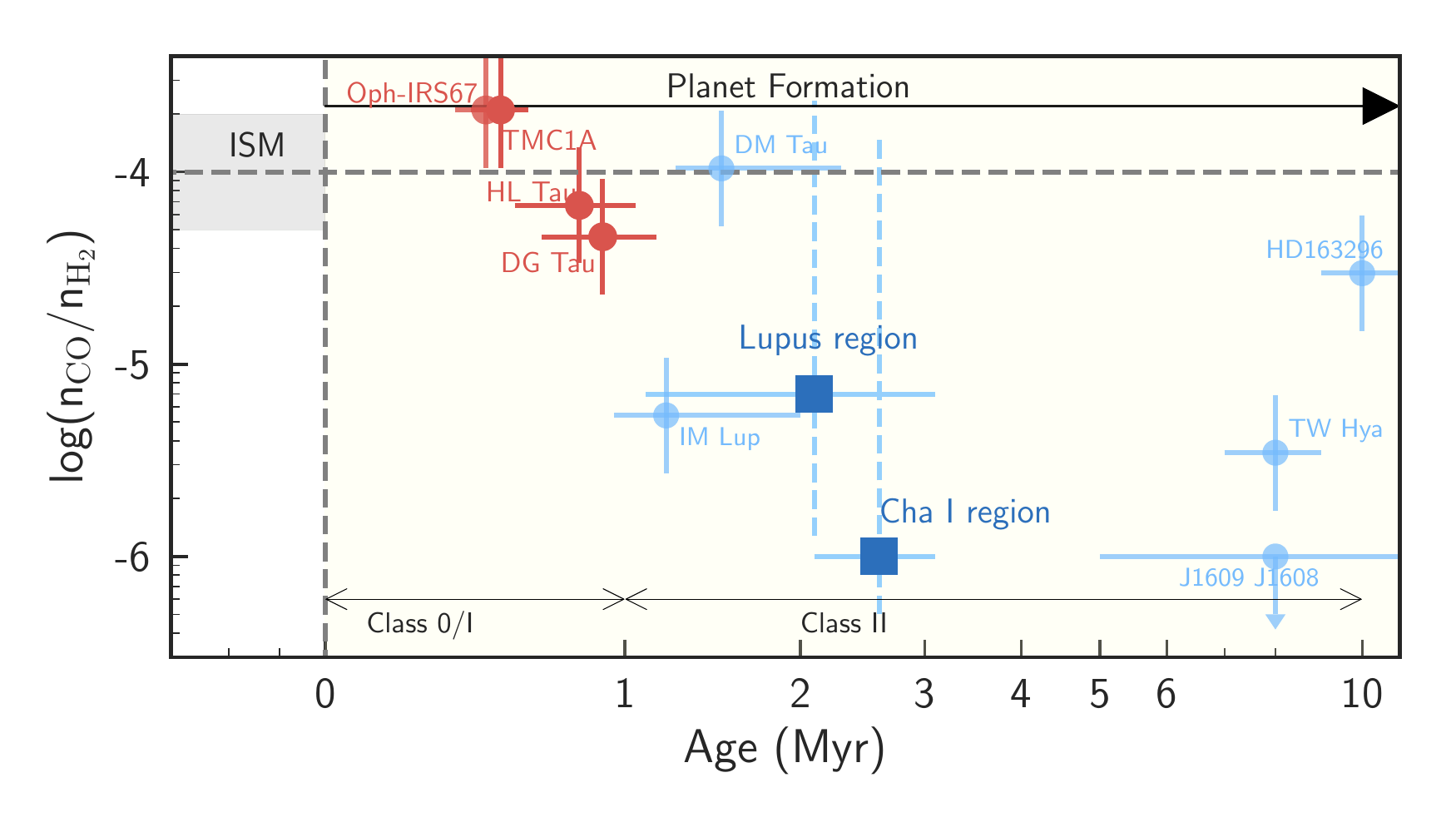}
\vspace{-0.5cm}
\caption{CO gas abundance measurements in protostellar and protoplanetary disks. The protostellar disks ($<$1\,Myr) and  protoplanetary disks (1-10\,Myr) are in red and blue, respectively. The filled circles are for individual disks and the squares for averaged values of star formation regions. For individual sources, the error bars indicate the expected uncertainties. For star formation regions, dash lines indicate the full ranges of CO abundances found in each region. The TMC1A, HL Tau, and DG Tau data are from this work, and the Oph-IRS67 from \citet{ArturdelaVillarmois18}. The DM Tau, IM Lup, HD 163296, and TW Hya data are from disk average CO abundances of \citet{Zhang19}. The J160900–190852 and J160823–193001 data point is from \citet{Anderson19}. The data of Lupus and Chamaeleon I region are from \citet{Ansdell16} and \citet{Long17}, respectively. The relative ages of the protostellar systems are based on their SED shapes \citep{robitaille07}. All data assumed a gas-to-dust mass ratio of 100. \label{fig:co_evolution}}
\end{figure*}

Chemical models with an ISM cosmic-ray rate generally require 1-3\,Myr to achieve a CO depletion by a factor of 10, which is too slow compared to the evolution timescale seen in Figure~\ref{fig:co_evolution}. If the cosmic-ray rate can reach 10$^{-16}$\,s$^{-1}$ or higher as predicted by \citet{Padovani16}, the depletion timescale can be shortened to less than 1\,Myr. 

The second type of mechanism is dust growth, which sequesters CO ice into the mid-plane. It requires that dust growth is ongoing beyond the mid-plane CO snowline \citep{Krijt18}. Protostellar disks are expected to be warmer than protoplanetary disks due to their higher accretion luminosity and surrounding envelopes \citep{Harsono15}. In our models, the CO snowlines are beyond the mm disks in HL Tau and DG Tau, and beyond 60\,au in TMC1A. If the disk is warmer than the CO condensation temperature in most of its mass region, little CO would be depleted even if significant dust growth already occurs. This is most clearly seen in the HL Tau disk. It has substantial dust growth into mm- to cm-sized particles, but its disk-averaged abundance of the CO gas is nearly the ISM value. 

In summary, the ISM level CO abundances measured in protostellar disks indicate that no more than half of the initial CO gas is depleted at the earliest stage. Dust growth at this stage cannot efficiently deplete the CO abundance due to the lack of CO freeze-out under the warm disk condition. Chemical processing can still be at work if sufficient ionization is provided, and thus may play an important role at the beginning of the CO depletion processes. But the known chemical paths of CO processing also become inefficient in regions warmer than 30\,K \citep{Bosman18}. 
To explain the fast evolution of the CO abundance, a coupling of physical and chemical processing may be necessary to accelerate the depletion processes.

\subsection{Comparing radii of gas and dust emissions of protostellar disks} \label{sec:disk_size}
The emitting radii of (sub)mm-continuum and CO lines of protoplanetary disks typically appear to differ by a factor of 2 or more \citep{andrews12,ansdell18}. This discrepancy 
is often attributed to fast radial drift of dust grains into the inner disk region \citep{Birnstiel14, Trapman19}. In our models of the three protostellar disks, we find that models with similar disk sizes of gas and mm-sized dust can better reproduce the observations than that with disk size separations. The lack of size difference can be attributed to neither a lack of dust growth nor a timescale problem. All three disks are known to have a significant amount of dust growth into mm or larger sizes, and dust growth models show that the dust and gas disks can differ by a factor of 2 within 10$^5$\,yr \citep{Birnstiel14}. The lack of size difference hints that the retention of mm-sized dust particles in protostellar disks can be different from protoplanetary disks. For example, \citet{Bae15} showed that protostellar infall can trigger the Rossby wave instability and form vortices that efficiently trap dust particles in prostellar disks. 

\subsection{CO as a gas mass tracer of protostellar disks}
The determination of masses of gas-rich disks around young stars is crucial for nearly all aspects of planet formation. So far, HD is the most robust gas mass tracer, but it has only been detected in three protoplanetary disks and is challenging to be used for protostellar disks. Low $J$-lines of HD ($J=1,2$) emit at far-IR wavelengths ($>$50$\mu$m), where the dust emission of the envelope is strong and optically thick. 

Given the relatively narrow range of CO gas abundances seen in protostellar disks, we propose that CO may still be a good mass tracer in warm protostellar disks. This is consistent with current theoretical expectation that CO processing through dust growth does not work in warm regions without CO gas freeze-out.
Future investigations of the usage of CO as gas mass tracer would require spatially resolved observations of CO isotopologue lines to evaluate the contribution of disk and envelope and observations of other chemical tracers to evaluate the chemical processing of CO. Observing \ccc~low $J$-lines in $\sim$10 warm protostellar disks with $<$50\,au resolution would be sufficient to test the robustness of CO as a gas mass tracer.

\acknowledgments
 This work is based on observations carried out under project number W17AZ001 and W18BN001 with the IRAM NOEMA Interferometer. IRAM is supported by INSU/CNRS (France), MPG (Germany) and IGN (Spain). We thank the IRAM staff member Jan-Martin Winters for assistance with observations and data calibrations. We thank the anonymous referee for insightful suggestions which helped us improve this work.
K.Z. and K.S. acknowledge the support of NASA through Hubble Fellowship grant HST-HF2-51401.001,and HST-HF2-51419.001 awarded by the Space Telescope Science Institute, which is operated by the Association of Universities for Research in Astronomy, Inc., for NASA, under contract NAS5-26555. EAB acknowledges support from NSF Grant\#1907653. 

\newpage
%

\facilities{NOEMA(PolyFix)}


\software{RADMC3D \citep{radmc3d},
          GILDAS \footnote{See http://www.iram.fr/IRAMFR/GILDAS for more information about the GILDAS softwares.}, astropy \citep{Astropy13}}

\bibliographystyle{aasjournal}
\bibliography{lib}{}



\end{document}